\documentclass[12pt]{article}
\RequirePackage{setspace}
    \AtBeginDocument{\setlength\abovedisplayskip{3pt}}
    \AtBeginDocument{\setlength\belowdisplayskip{3pt}}
\usepackage{sectsty}
\sectionfont{\normalsize}
\usepackage{appendix}
\usepackage{float}
\usepackage{url}
\usepackage{caption}
\usepackage[round]{natbib}

\usepackage{amsthm}
\usepackage{amsfonts}
\usepackage{bbm}
\usepackage{graphicx}
\usepackage{sectsty}
\sectionfont{\large}
\usepackage{appendix}
\usepackage{float}
\usepackage{url}

\newcommand{\be}{\begin{eqnarray}}
\newcommand{\ee}{\end{eqnarray}}

\newcommand{\captionfonts}{\footnotesize}
\makeatletter  
\long\def\@makecaption#1#2{%
  \vskip\abovecaptionskip
  \sbox\@tempboxa{{\captionfonts #1: #2}}%
  \ifdim \wd\@tempboxa >\hsize
    {\captionfonts #1: #2\par}
  \else
    \hbox to\hsize{\hfil\box\@tempboxa\hfil}%
  \fi
  \vskip\belowcaptionskip} 
\makeatother   
\begin{document}
\title{Quantum structure in competing lizard communities\footnote{\it This paper is dedicated to the memory of our dear colleague and friend Prof. Bart D'Hooghe, who recently passed away, after collaborating with us for several years and contributing to the first draft of this manuscript.}}
\author{\normalsize Diederik Aerts$^1$, Jan Broekaert$^1$, Marek Czachor$^{1,2}$, Maciej Kuna$^3$, Barry Sinervo$^4$ and Sandro Sozzo$^1$ \\ \\
        \small\itshape \vspace{-0.1 cm}
        $^1$ Center Leo Apostel for Interdisciplinary Studies \\
       \vspace{-0.1 cm} \small\itshape
         Department of Mathematics and Department of Psychology \\
         \vspace{-0.1 cm} \small\itshape
         Brussels Free University, Brussels, Belgium \\
       \vspace{-0.1 cm} \small
        Emails: \url{diraerts@vub.ac.be, jbroekae@vub.ac.be, ssozzo@vub.ac.be} \vspace{0.2 cm} \\ 
        \small\itshape \vspace{-0.1 cm}
        $^2$ Department of Theoretical Physics and Quantum Information \\
       \vspace{-0.1 cm} \small\itshape
         Politechnika Gda\'nska, Gda\'nsk, Poland \\
       \vspace{-0.1 cm} \small
        Email: \url{mczachor@pg.gda.pl} \vspace{0.2 cm} \\ 
        \small\itshape \vspace{-0.1 cm}
        $^3$ Department of Probability and Biomathematics \\
       \vspace{-0.1 cm} \small\itshape
         Politechnika Gda\'nska, Gda\'nsk, Poland \\
       \vspace{-0.1 cm} \small
        Email: \url{maciek@mif.pg.gda.pl} \vspace{0.2 cm} \\ 
        \small\itshape \vspace{-0.1 cm}
         $^4$ Department of Biology and Evolutionary Biology \\
       \vspace{-0.1 cm} \small\itshape
         University of California, Santa Cruz, California, USA \\
       \vspace{-0.1 cm} \small
        Email: \url{lizardrps@gmail.com} \vspace{0.2 cm} \\
         }
\date{}
\maketitle
\vspace{-1 cm}
\begin{abstract}
\noindent
Almost two decades of research on applications of the mathematical formalism of quantum theory as a modeling tool in domains different from the micro-world has given rise to many successful applications in situations related to human behavior and thought, more specifically in cognitive processes of decision-making and the ways concepts are combined into sentences. In this article, we extend this approach to animal behavior, showing that an analysis of an interactive situation involving a mating competition between certain lizard morphs allows to identify a quantum theoretic structure. More in particular, we show that when this lizard competition is analyzed structurally in the light of a compound entity consisting of subentities, the contextuality provided by the presence of an underlying rock-paper-scissors cyclic dynamics leads to a violation of Bell's inequality, which means it is of a non-classical type. We work out an explicit quantum-mechanical representation in Hilbert space for the lizard situation and show that it faithfully models a set of experimental data collected on three throat-colored morphs of a specific lizard species. Furthermore, we investigate the Hilbert space modeling, and show that the states describing the lizard competitions contain entanglement for each one of the considered confrontations of lizards with different competing strategies, which renders it no longer possible to interpret these states of the competing lizards as compositions of states of the individual lizards.
\end{abstract}

\begin{quotation}
\end{quotation}

\section{Introduction\label{intro}}
This article looks into the challenging question of whether quantum structures are present in aspects of animal behavior. More specifically, we discuss an example that reveals contextuality and the appearance of entanglement in a proposed quantum theoretic model for the mating competition of three male morphs of \emph{Uta stansburiana} lizards.

The first step leading to the result we put forward in the present article was related to our study of biological evolution based on a specific situation involving the rock-paper-scissors (RPS) game as an example. We observed \citep{evolution2011} that when the RPS game was regarded as a coincidence experiment, it allowed for violation of Bell's inequality \citep{bell1964}.

The presence of contextuality in a situation of a compound entity consisting of two subentities essentially means that what happens with one of the subentities affects the behavior of the other subentity, which as a general situation is quite common. Contextuality can hence readily be identified in the case of the RPS dynamics, when the two players involved in the interaction are looked upon as a compound entity comprising two subentities. Indeed, whether one of the players wins or loses depends essentially on what the other player does.

It has been shown in the foundations of quantum theory that if this contextuality -- in addition to its readily identifiable effect of one subentity functioning as a context for the other subentity -- leads to a violation of Bell's inequality, this is indicative of the presence of a special type of contextuality which cannot be modeled classically and which, when modeled quantum-mechanically, is expressed by the appearance of entanglement in the state of the compound entity \citep{AccardiFedullo1982,Aerts1986,Pitowsky1989}. In the following we will refer to this type of `Bell's inequality violating contextuality' as `non-classical contextuality'. The presence of entanglement in the state of the compound entity means that this state cannot be described any longer as the composition of two states, where each one is a state of one of the subentities. The compound entity entails a new type of difficulty when attempted to be interpreted as two subentities, which is the fingerprint of the presence of quantum structure for any situation of compoundness. This is why finding Bell's inequality to be violated by the ideal RPS game to us was a straightforward reason to investigate whether a quantum structure was involved in this RPS dynamics.

The relevance of this insight to biology only became clear when some of us learned that the RPS game had been used as a modeling scheme for specific types of dynamical situations in population ecology, including the famous `paradox of the plankton', where it is referred to as cyclic or multiple competition \citep{HW99,Huismanetal2001,Schippers,Laird,AllesinaLevine2011}. Also, situations of competing lizard species were studied intensively by considering cyclic competition as a fundamental aspect of their dynamics. More specifically, one of us discovered an RPS strategy in the mating behavior of the side-blotched lizard species \emph{Uta stansburiana} \citep{S96,SL96}. It was found that males, having either orange, blue or yellow throats, follow heritable mating strategies. As in the RPS game, where scissors cut paper, rock crushes scissors, and paper wraps rock, the three-morph mating system is such that the wide-ranging ultradominant strategy of orange males is defeated by the sneaker strategy of yellow males, which is in turn defeated by the mate-guarding strategy of blue males. The orange strategy defeats the blue strategy, to complete the dynamic cycle. This `lizard game' presents a stable pattern in the replicator dynamics where the dynamical system follows closed orbits around a mixed strategy Nash equilibrium \citep{S96,SL96,Sinervo2001,SinervoetalPNAS2006,Sinervoetal2007}. And indeed, if we regard two competing lizard morphs as a compound entity of two individual lizard morphs, we can recognize the same type of contextuality that we identified for the ideal RPS game; whether one of the lizards in competition will impregnate a female depends essentially on the color of the other lizard. Additionally, Bell's inequality is violated also for the lizard morphs, as we will explicitly show in Section \ref{contextuality}, investigating in detail the contextuality that is apparent in the lizard competition. This explains our motivation to build a quantum-theoretic model for this lizard ecosystem.

Identification of quantum structure in the lizard dynamics can be seen as an example of the use of the mathematical formalism of quantum theory as a modeling instrument in domains different from the micro-world. This approach has led to interesting results in recent years and is now an active and emergent research field in itself. In cognitive science (concept theory and decision theory), in economics (finance and behavioral economics), and in computer science (semantic theories, information retrieval, and artificial intelligence), several situations have been identified where application of classical structures is problematic, whereas modeling based on quantum structures is successful \citep{aertsaerts1995,rijsb2004,aertsczachor2004,aertsgabora2005ab,busemeyeretal2006,pothosbusemeyer2009,bruza2009,aerts2009a,lambert2009,khrennikov_ubi,busemeyeretal2011,trueblood2011,ags12,busemeyerbruza2012,pothosbusemeyer2012,abgs12}. 
An important point to be made for the above-mentioned approaches is that it is not the presence of microscopic quantum processes that is considered to be at work to give rise to the appearance of quantum structure in these different domains. Rather the situation is such, that it is possible to identify in these domains some typical quantum features, such as the quantum-type of contextuality and entanglement, and it is these features themselves that give rise to the presence of quantum structure. We will identify this type of quantum structure for the lizard ecosystem. It is interesting to mention in this respect, that in a comparable way such quantum structure has been found to be quite systematically present in human cognition, in the processes of decision-making \citep{aertsaerts1995,busemeyeretal2006,pothosbusemeyer2009,busemeyeretal2011,trueblood2011,busemeyerbruza2012,pothosbusemeyer2012}, and in the dynamics of how humans use and combine concepts \citep{aertsgabora2005ab,aerts2009a,ags12,abgs12}.

In our investigation of the lizard ecosystem we construct an explicit quantum-theoretic representation in a complex Hilbert space of the underlying RPS-like dynamics that gives rise to the cyclic pattern of frequencies in the population identified experimentally. To accomplish this, we make use of the specific rules of the quantum formalism to calculate the probabilities in this underlying RPS-like dynamics in a way that allows to faithfully represent the experimental data gathered by one of us on the population frequencies over the last two decades. In Sections \ref{lizard_game} and \ref{outcomeprobabilities}, we introduce our lizard system and explain the main aspects of our approach and modeling of the underlying RPS-like dynamics. We analyze how contextuality is one of its essential features. The latter notion is analyzed in detail with respect to the lizard competition in Section \ref{contextuality}. In Section \ref{hilbertspacemodel}, we put forward the notions of the quantum-mechanical formalism that are needed in our paper, and in parallel we work out a Hilbert space model for the RPS-type lizard game. We show that the self-adjoint operators representing the confrontation events (called `measurements' in quantum jargon) in the lizard competition do not commute, which means that the probability structure connected to them is non-classical. In Section \ref{compoundness} we analyze the `lizard morphs situation' explicitly from the perspective of a compound entity consisting of a subentities situation, a situation well-known and studied in quantum theory, and we show that, following such a quantum analysis of compoundness, this lizard morphs situation involves entanglement in its states for each of the considered measurements. The problem with a Kolmogorovian probability model for the lizard game is analyzed in Section \ref{kolmogorov}. Finally, in our conclusions of Section \ref{conclusions}, we put forward ideas for future investigation.
The general result obtained supports intuitions that dynamical systems based on non-Kolmogorovian probability may provide a fruitful conceptual framework for real-life interactions of populations \citep{ACKS_Ecol_Mod}.

\section{The RPS-type nature of the lizard dynamics\label{lizard_game}}
Before we provide proof of a quantum-like dynamical structure underlying the competing morphs of the lizard {\it Uta stansburiana}, we briefly sketch some game-theoretic aspects of population ecology.

Species competition can be reformulated in terms of evolutionary game dynamics describing how the frequencies of strategies within a population change in time, according to their success. Game theory typically deals with an individual (\emph{player}) who is engaged in a given interaction (\emph{game}) with other players and can decide between different options (\emph{strategies}). Depending on the strategies of a player and its co-players a \emph{payoff} is realized, and the possible maximization of this payoff is one of the fundamental aspects of game-theory. Evolutionary game dynamics thus deals with populations of players programmed -- genetically or possibly also induced by the environment -- to use the same strategy. Strategies with high payoff will spread within the population, where the payoffs depend on the actions of the co-players and hence on the frequencies of the strategies within the population. In classical evolutionary game theory, one typically assumes that the elements of the pay-off matrix are time invariant and evolution of the system takes place as frequency-dependent fitness changes, thereby changing the relative success and the probability of encountering each strategy over time. In evolutionary biology, the strategies can be identified with morphs, and many species exhibit color polymorphisms associated with alternative male reproductive strategies \citep{SL96,Sinervoetal2007,SinervoCalsbeek2006}. The prevalence of multiple morphs is a challenge to evolutionary theory because a single strategy should prevail unless morphs have exactly the same fitness or a fitness advantage when rare. One of us has shown in several papers that the three color morphs of side-blotched lizards, \emph{Uta stansburiana}, follow an underlying RPS-like dynamics \citep{SL96,Sinervo2001,SinervoetalPNAS2006,SinervoCalsbeek2006,BleaySinervo2007}. More precisely, males have either orange ($o$), blue ($b$) or yellow ($y$) throats and each type follows a fixed mating strategy, as follows:

(i) Orange-throated males are strongest and do not form strong pair bonds; instead, they fight blue-throated males for their females. Yellow-throated males, however, manage to copulate with females in the orange male harems. The large size and aggression is caused by high testosterone production \citep{Millsetal2008}.

(ii) Blue-throated males are smaller in size and form strong pair bonds. While they are outcompeted by orange-throated males, they can defend against yellow-throated ones via co-operation with other blue-throated neighbors. Because blue-throated males produce less testosterone, they are not as strong as the orange-throated males, but it gives them the advantage of being less aggressive and able to form strong pair bonds, and also engage in territorial co-operation with neighboring blue-throated males \citep{SinervoetalPNAS2006}.

(iii) Yellow-throated males are smallest, and their coloration mimics females. This enables them to approach females in the harems of orange-throated males and mate when the latter are distracted. This is less likely to work with a female that has bonded with a blue-throated male, and by virtue of his vigilant co-operative blue-throated male partner.

Points (i)-(iii) can be summarized as ``$o$ beats $b$, $b$ beats $y$, and $y$ beats $o$'', which is similar to the RPS rules. Therefore $o$ and $y$ provide contexts for $b$, and in turn, $b$ and $y$ provide contexts for $o$, and finally $o$ and $b$ provide contexts for $y$.
Thus, the interaction of the `RPS lizards' exposes a deeper underlying contextuality beyond the individual players, since their strategy is unchangeably fixed by their color.

Fundamentally it is the structure of the underlying RPS-like dynamics that entails the quantum structure of the model we will construct. For an ideal RPS-situation, where paper beats rock, scissors beats paper, and rock beats scissors, no probabilities are involved but certitudes (probabilities equal to 1 or 0). We will see in the following that, even for this ideal RPS situation, Bell-type inequalities are violated, which shows that also in this deterministic limit case, contextuality is present.
It is contextuality which gives rise to non-classical quantum-like structure tested by the Bell-type inequalities. Of course, the real-world situation, with male lizards confronting each other in competition for a female, does not reflect the ideal RPS-situation.
The experimental data (next section) shows that outcome probabilities not equal to 1 or 0 are valid for the real-world situation. This means that the contextuality -- which is deterministic for the true RPS-game -- is probabilistic in the real-world case of the lizards.

We must therefore first calculate, for the underlying RPS-like situation, the outcome probabilities of male lizards of different colors winning or losing a mutual competition for females. Although these outcome probabilities are at the origin of the measured cyclic fluctuations in the frequency data collected in lizard experiments, they cannot be measured directly. In the next section we explain how we calculate the outcome probabilities starting from data collected by \cite{BleaySinervo2007}. The first step in our aim is to (i) prove that the RPS-like outcome probabilities violate Bell's inequalities, and hence cannot be represented within a Kolmogorovian probability model and, (ii) build a Hilbert space model that does represent these probabilities quantum-theoretically, and show explicitly how the considered measurements correspond to non-commuting observables within this Hilbert space representation.

\section{Calculating the RPS-type outcome probabilities \label{outcomeprobabilities}}

In \cite{BleaySinervo2007}, the mutual confrontations of the different lizard morphs are described as a cyclic RPS-like dynamics: {\it orange beats blue}, {\it blue beats yellow} and {\it yellow beats orange}. So we can choose to identify {\it orange} with {\it rock}, {\it blue} with {\it scissors} and {\it yellow} with {\it paper}.

In order to identify the RPS-like scheme quantitatively we consider their experiments more closely. The relevant data (\cite{BleaySinervo2007}, Fig. 1) concerns `male fitness', measured by counting the proportion of clutch sired by the different male morphs, orange, yellow, or blue in a frequency-controlled female environment. The experiments were performed in three variations, with each variation specifically controlling the `male morph frequency within a female's social neighbourhood'. In particular, one variation controlled the orange male morph frequency within a female's social neighbourhood, the second variation, the yellow male morph frequency, and the third, the blue male morph frequency.

A first consideration of these data shows how the `outcome probabilities' are contained in the morph color frequency of the female's clutch measured in the experiment. E.g. in the variant with controlled frequency of orange morphs in a female's social neighbourhood, the all-orange female environment leads to the clutch proportions of 0.28 orange hatchlings, 0.53 yellow hatchlings, and 0.19 blue hatchlings. Thus, socially surrounded by all-orange morphs, the yellow `sneakers' manage to sire more hatchlings than the orange ones. The blue morphs are less successful in this situation. These clutch color proportions therefore express a weighted mean of outcome probabilities of mutual morph color competitions.

In order to extract outcome probabilities from these clutch color proportions, we put forward the following hypotheses concerning the competition for a female between a particular color morph and another particular color morph, where the female's clutch is sired by one of them.

(i) Only one of the competing males sires the clutch of the female, and since it is color proportion we measure, only one of the male morph colors `wins' and the other morph color then `loses', with respect to this measurement of fertility. We thus define `win' and `lose' in the following way. A specific morph color `wins' in a competition with another morph in case its color is transferred to the hatchlings being born in the clutch of the female they compete for. The morphs is defined to `lose' in case it does not win.

(ii) We specify in a way compatible with our general definition of `win' and `lose' what happens in case of a `draw'. With the win-lose definition we just introduced, in the case of two morphs of the same color competing, a hatchling is sired by their common color and as a consequence they both win.

(iii) The competition situation is symmetric. We will suppose that only the confrontation of the two different strategies corresponding to the two morph colors determines the dynamics and hence the result of winning or losing.

(iv) Since even if in a female's social neighbourhood males of only one morph color are maximally present, all morph colors still appear among the hatchlings, the underlying dynamics cannot be that of the ideal RPS-game. Thus color competition outcome probabilities can be different from both 1 and 0.

We introduce the following notation to be able to express the content of hypotheses (i), (ii), (iii) and (iv). We denote by
\begin{equation} \label{jointprobabilities}
p(ab)_{11} \quad p(ab)_{12} \quad p(ab)_{21} \quad p(ab)_{22} \quad
\end{equation}
the probabilities that for two morphs with colors $a$ and $b$ confronting each other, there is a situation of `win, win', this is $p(ab)_{11}$, of `win, lose', this is $p(ab)_{12}$, of `lose, win', this is $p(ab)_{21}$, and of `lose, lose', this is $p(ab)_{22}$. Hence, the `1' or `2' in the subscript slots signify `win' and `lose'.

We will, for $a$ and $b$ interchangeably, allow the colors of the morphs to be used, with letters $o$, $y$ and $b$. Since the four outcomes $\{11, 12, 21, 22\}$ exhaust all possibilities, we have:
\begin{equation}
\sum_{ij=1}^2p(ab)_{ij}=1 \label{normprob}
\end{equation}

The above-mentioned hypotheses can now be expressed mathematically by means of these probabilities.
\begin{eqnarray}
&&p(ab)_{11}=p(ab)_{22}=0 \quad p(ab)_{12}+p(ab)_{21}= 1 \quad {\rm if} \quad a \not=b \label{antisymmetry} \\
&&p(aa)_{11}=1 ,\quad p(aa)_{12}=p(aa)_{21}=p(aa)_{22}=0 \label{drawprobability}\\
&&p(ab)_{ij}=p(ba)_{ji} \label{symmetry} \\
&& 0\leq p(ab)_{ij} \leq 1 \label{naturalprobabilities}
\end{eqnarray}
Now we complete the model with the observation we previously made in the experiment \citep{BleaySinervo2007}, and its first variation and specifically in its all-orange female social neighbourhood. Let us denote by $w_o(o)=0.35$, $w_o(y)=0.41$ and $w_o(b)=0.24$ the proportions of orange, yellow and blue hatchlings found in the female's clutch.
Each of these color appearances is the consequence of competitions taking place in the all-orange environment. And competitions can in principle be of six different types, orange-orange, orange-yellow, orange-blue, yellow-yellow, yellow-blue and blue-blue. In a given situation of a female's social neighbourhood the color proportions in the clutch are therefore determined completely by (i) the fraction of each type of the six possible confrontations taking place, and (ii) the outcome probabilities for each of the confrontations, because indeed, due to (\ref{antisymmetry}) and (\ref{drawprobability}), each one of such confrontationd leads to one unique color in the clutch, and the probability of this happening is given by the outcome probability.

Since we do not know the relative importance of each type of confrontation, we will assign a number $P_c(ab)$ to the fraction of confrontations of colors $a$ and $b$ taking place in the female social neighbourhood of color $c$. Since color confrontations are symmetric, these assigned weights satisfy $P_c(ab)=P_c(ba)$. 
Since $P_c(ab)$ represent fractions, we have, for an arbitrary $c$, the following normalisation:
\begin{eqnarray}
P_c (ob) + P_c (oy) + P_c (oo) + P_c (yb) + P_c (yy) + P_c (bb) &=& 1\label{bicolorcompetition}
\end{eqnarray}
Each $P_c(ab)$ represents the fraction of confrontations $(ab)$ that, due to (\ref{antisymmetry}) and (\ref{drawprobability}), can lead to offspring with either color $a$ or $b$. 
This means that, for $a\not=b$, and applying (\ref{antisymmetry}), $P_c(ab)p(ab)_{12}$ is the proportion of $a$-colored offspring in the clutch, for a female environment of color $c$ -- since $p(ab)_{12}$ is the probability that $a$ wins over $b$
 -- and $P_c(ab)p(ab)_{21}$ is the proportion of $b$-colored offspring in the clutch, for a female environment with color $c$, since $p(ab)_{21}$ is the probability that $b$ wins over $a$.
Finally applying (\ref{drawprobability}), we find $P_c(aa)p(aa)_{11}=P_c(aa)$ is the proportion of $a$-colored offspring in the clutch, for a female environment of color $c$.

We expect in environment $c$ most of the confrontations to be of the type $(ca)$, with $a$ one of the three colors, and hence confrontations of the type $(ab)$, with $a\not=c$, and $b\not=c$, to be minimal, possibly negligible. The derivation of the outcome probabilities from the data in \cite{BleaySinervo2007} that we present in the following, is general enough to take into account this asymmetry due to the focus on the social environment color of the female, as will become clear in the following. 
Besides the coloured proportions of the clutch in orange social environment $w_o(a)$, we introduce now the notation $w_c(a)$ expressing the proportions of $a$-colored hatchlings found in the female's clutch for a social environment of color $c$ in general. Since all color-specific proportions constitute the full clutch, we have for arbitrary $c$,
\begin{equation} 
w_c(o)+w_c(y)+w_c(b)=1
\end{equation}
and the values of all nine $w_c(a)$ were determined experimentally in \cite{BleaySinervo2007}. We have introduced now all necessary elements to derive the equations for calculating the outcome probabilities from the fractions of the colors in the clutch. For an arbitrary $c$, representing the color of the female social environment, we have
\begin{eqnarray} \label{orange}
P_c(oo)p(oo)_{11}+P_c(oy)p(oy)_{12}+P_c(ob)p(ob)_{12}=w_c(o) \\ \label{yellow}
P_c(yy)p(yy)_{11}+P_c(oy)p(oy)_{21}+P_c(yb)p(yb)_{12}=w_c(y) \\ \label{blue}
P_c(bb)p(bb)_{11}+P_c(ob)p(ob)_{21}+P_c(yb)p(yb)_{21}=w_c(b)
\end{eqnarray}
Let us interpret these equations, for example, for an orange social environment of the female. 
There are three of the six possible confrontations that can give rise to orange offspring, they are $(oo)$, $(oy)$, and $(ob)$. Each time, however,  they will contribute to orange offspring only if `orange wins'. The outcome probabilities expressing these events are $p(oo)_{11}=1$, $p(oy)_{12}$ and $p(ob)_{12}$. This is the content of (\ref{orange}) for $c$ being orange. There are also three of the six possible confrontations that can give rise to yellow offspring, they are $(yy)$, $(oy)$, and $(yb)$. Each time, however, they will contribute to yellow offspring only if `yellow wins'. The outcome probabilities expressing these events are $p(yy)_{11}=1$, $p(oy)_{21}$ and $p(yb)_{12}$. This is the content of (\ref{yellow}) for $c$ being orange. There are again three of the six possible confrontations that can give rise to blue offspring, they are $(bb)$, $(ob)$, and $(yb)$. Each time, however, they will contribute to blue offspring only if `blue wins'. The outcome probabilities expressing these events are $p(bb)_{11}=1$, $p(ob)_{21}$ and $p(yb)_{21}$. This is the content of (\ref{blue}) for $c$ being orange. In a straightforward generalization, the six additional equations, three for $c$ being yellow, and three for $c$ being blue, are interpreted analogously.

We must now solve these nine equations. The right-hand side elements of the equations are experimentally determined, and more specifically we extract from \cite{BleaySinervo2007}, Figure 1, the following values
 \begin{eqnarray}
&w_o(o)= 0.28 \quad\quad\quad    w_o(y)  =  0.53 \quad\quad\quad   w_o(b)   = 0.19& \\
 &w_y(o)= 0.15  \quad\quad\quad    w_y(y)  =  0.30 \quad\quad\quad   w_y(b)   = 0.55& \\
 &w_b(o)= 0.54 \quad\quad\quad    w_b(y)  = 0.14 \quad\quad\quad   w_b(b)   = 0.32&  \label{solution03}\end{eqnarray}
Also the three trivial outcome probabilities $p(oo)_{11} = p(yy)_{11}= p(bb)_{11}= 1$ are known. Only three of the outcome probabilities $p(ab)_{ij}$ are independent due to symmetries and closure, (\ref{antisymmetry}), (\ref{symmetry}), and fifteen of the fractions of confrontations $P_c(a,b)$ are independent due to symmetry and closure.\\
While this system of equations does not lead to a unique solution, one can check as proof of concept that the following solutions solve the system of equations:
 \begin{eqnarray} \label{solutionob}
&p(oy)_{21} = 0.88  \quad\quad\quad p(oy)_{12} = 0.12 & \\ \label{solutionoy}
&p(yb)_{21}  = 0.82  \quad\quad\quad p(yb)_{12} = 0.18 &\\ \label{solutionyb}
&p(bo)_{21}  = 0.72   \quad\quad\quad p(bo)_{12}  = 0.28 &
\end{eqnarray}
with
\begin{eqnarray}
&P_o(oo)=  0.10 \quad\quad\quad    P_o(oy)  = 0.49  \quad\quad\quad   P_o(ob)   = 0.17 & \\
 &P_o(yy)= 0.08   \quad\quad\quad P_o(yb)= 0.08 \quad\quad\quad P_o(bb)= 0.08 & \\
 &P_y(oo)= 0.08  \quad\quad\quad    P_y(oy)  = 0.11 \quad\quad\quad   P_y(ob)   =0.08 & \\
 &P_y(yy)=  0.11  \quad\quad\quad P_y(yb)= 0.55   \quad\quad\quad P_y(bb)= 0.07 &\\
 &P_b(oo)= 0.08 \quad\quad\quad    P_b(oy)  = 0.07  \quad\quad\quad   P_b(ob)   = 0.63 & \\
 &P_b(yy)= 0.06  \quad\quad\quad P_b(yb)=  0.08  \quad\quad\quad P_b(bb)= 0.08&
 \end{eqnarray}
This solution was generated using a simplex algorithm and can be steered by feeding initial values to the parameters. The present solution should not necessarily reflect a true natural configuration, but only one possibility steered for small weights to confrontations between two morph colors in a third-color social environment.

The model thus shows that the outcome probabilities obtained from the experimental data confirm RPS-like probabilities: $p(oy)_{21}$,  $ p(yb)_{21}$ and $(bo)_{21}$ are all inclined to 1 for the RPS-`win', while $p(oy)_{12} $, $ p(yb)_{12} $ and $p(bo)_{12}$ are inclined to 0 for the typical RPS-`lose'.

\section{Contextuality, the violation of Bell inequalities and of the marginal law\label{contextuality}}
The analysis in the foregoing section reveals the essential aspect of the underlying RPS-type dynamics giving rise to the cyclic permutations in the population densities, when the outcome probabilities are to be modeled. It consists in considering the interaction situation of two lizards competing for a female from the perspective of a compound entity (the two interacting lizards), consisting of two subentities (each of the two lizards apart), and joint measurements to be performed on the compound entity (the competing strategies of the two lizards), resulting in outcomes that can be interpreted for each of the subentities apart (both lizards can win or lose, defined by `transferring its color to the offspring', and hence the `win' or `lose' is defined for each of the lizards apart). This situation of `compound entity' and `joint measurements performed on the compound entity, with outcomes interpretable as outcomes for the subentities', was investigated in great detail in quantum theory, and will be used to analyze the lizards configuration \citep{1asQI2013,2asQI2013,asIQSA2012}.

The Clauser-Horne-Shimony-Holt (CHSH) variant of Bell's inequalities is defined in physics by means of the `expectation values' of the joint measurements. These expectation values are nothing but the weighted average value of an outcome, and the outcome values themselves are typically \emph{set} to $+1$  or $-1$. We will first briefly explain the content of the CHSH inequality and subsequently relate it to the present biological context.

Regarding the general content of the CHSH inequality, one begins by considering a compound entity $S$ comprising two subentities $S_1$ and $S_2$ and prepared in a given state. This is followed by simultaneous measurements of the observables $a$ and $b$, each with possible outcomes $\pm 1$, on $S_1$ and $S_2$, respectively. The statistics of outcomes are collected and one calculates the expectation values $E(ab)=p(ab)_{11}+p(ab)_{-1-1}-p(ab)_{-1+1}-p(ab)_{-1+1}$. Here, $p(ab)_{ij}$ (with $i,j=\pm 1$) is the probability of obtaining the pair $(i,j)$ when measuring $ab$, the joint measurement of $a$ on $S_1$ and $b$ on $S_2$, on the joint system $S$. This procedure is repeated for the pair of measurements $(a,b')$, $(a',b)$ and $(a',b')$. It is possible to prove that the collected joint probabilities can be cast into a global classical Kolmogorovian probability space if and only if the following `Bell's inequality' is satisfied
\begin{equation}
\label{chshphysics}
-2 \le E(ab)-E(ab')-E(a'b)+E(a'b')\le 2
\end{equation}
It has been shown for micro-physical entities described by quantum theory that if the initial state of the compound entity and the measurements are properly chosen, the Bell's inequality in (\ref{chshphysics}) is violated, which entails in particular that quantum probabilities cannot be recovered in a Kolmogorovian framework, i.e. they are non-Kolmogorovian. A case in which this violation occurs is the case where the compound entity is prepared in a suitable `entangled state', i.e. a state that cannot be written as a product of a state of $S_1$ and a state of $S_2$ (see Section \ref{compoundness}). One refers to such a situation of violation due to the presence of an entangled state as entanglement, and identifies entanglement as one of the most important non-classical aspects of quantum theory. 
Obviously, since the CSHS-inequality is merely a statistical tool, it can be applied to any kind of entities, not necessarily pertaining to particles of physics. It is straightforward to adapt it to the lizard ecosystem, by suitably introducing states, joint measurements and probabilities of outcomes, and it is remarkable that a simple calculation suffices to obtain the violation of the inequalities.

To this end, we proceed as follows. In the foregoing section we introduced the joint probabilities for the joint measurements. For the case of two interacting lizards with colors $a$ and $b$ considered as a compound entity of the two subentities which are the two individual lizards, they are given in (\ref{jointprobabilities}). 
In line with its definition in probability theory, we assign the value `+1' to the outcome if a confrontation of two morphs is of the symmetric type `win, win', or `lose, lose'. And we assign the value `-1' to the outcome if a confrontation of two morphs is of the mixed type `win, lose', or `lose, win'. This means that the expectation value for such a joint measurement is
\begin{equation}
E(ab)=p(ab)_{11}-p(ab)_{12}-p(ab)_{21}+p(ab)_{22}
\end{equation}  
We can then use (\ref{chshphysics}) above, where $a$, $b$, $a'$, and $b'$ are different colors of morphs.

For two colors $a$ and $b$ we have
\begin{eqnarray}
&&E(ab)=-1 \quad {\rm if} \quad a\not=b \label{anticorrelation}\\
&&E(aa)=+1 \label{correlation}
\end{eqnarray} 
Indeed, suppose that $a\not=b$, then from (\ref{antisymmetry}) we have $p(ab)_{11}=p(ab)_{22}=0$ and $p(ab)_{12}+p(ab)_{21}=+1$. This means that
\begin{eqnarray}
E(ab)=p(ab)_{11}-p(ab)_{12}-p(ab)_{21}+p(ab)_{22}=-1
\end{eqnarray}
Using (\ref{drawprobability}), and hence $p(aa)_{11}=1$, and $p(aa)_{12}=p(aa)_{21}=p(aa)_{22}=0$, it follows that
\begin{eqnarray}
E(aa)=p(aa)_{11}-p(aa)_{12}-p(aa)_{21}+p(aa)_{22}=+1
\end{eqnarray}
To violate the CHSH variant of Bell's inequality, consider the following colors. We take $a$ to be orange, $b$ also to be orange, $a'$ to be yellow and $b'$ to be blue. We then have $E(oo)=+1$ and $E(ob)=E(yo)=E(yb)=-1$. This gives
\begin{equation}
E(oo)-E(ob)-E(yo)-E(yb)=+4
\end{equation}
hence a maximal violation with value +4 of the inequality.

Instead of analyzing exactly what the violation of the inequalities (\ref{chshphysics}) implies for the lizards system, we will in the remainder of this paper explicitly construct the entangled states for the lizards system, which are the real source for the violation of these inequalities. Indeed, it should be noted that in effect the violation of the CHSH inequality with the expectation values captures only a restricted aspect of the source of this entanglement. This is also the case in micro-physics, and it explains why many entangled states can be realized that do \emph{not} violate the inequalities. The inequalities are only violated for specific and rather `extreme' states of entanglement. This means that it provides deep structural insights neither here nor in micro-physics to focus on the violation itself, such structural insights being linked to the source of the violation, which is entanglement. However, it can be shown that if the inequalities are violated, the presence of entanglement must be the cause of this violation.

Additionally, in physics the connection has been investigated (\cite{AccardiFedullo1982,Aerts1986,Pitowsky1989}) between the Bell inequality, its being satisfied or violated, and the possibility to fit the joint probability used for its expectation values in a general Kolmogorovian model. More precisely, it has been proven that if the Bell-inequality is violated, the joint probabilities used for its expectation values cannot be fit into a global Kolmogorovian model for the considered situation -- this will be investigated in detail in Section \ref{kolmogorov}.

Let us identify contextuality in the specific case of the lizards. As we can see here, the question of which lizard wins or loses depends essentially on the other lizard. This is true of the RPS game and it is true of the ideal lizard dynamics. None of the players can influence the outcome of a specific strategy in any way, because it wholly depends on the other player's strategy.
This type of direct outcome dependence, within a situation of a compound entity, and outcomes and joint measurements identifiable for the subentities, is known as `contextuality'. Of course, the above situation can also be simply a case of classical contextuality. However, since we have shown that Bell-type inequalities like CHSH (\ref{chshphysics}) are violated for the lizard situation, and, as we remarked already, this violation proves the existence of entanglement, we can state that also the contextuality identified in the lizard situation is of a non-classical type, linked to the presence of entanglement. Additionally, in Section \ref{hilbertspacemodel} below we will show that the lizard contextuality is related in a direct way  to non-commutativity of measurements, which makes it non-classical also with respect to this aspect of non-classicality, i.e. non-commutativity. How it is connected to entanglement itself will be investigated in detail in Section \ref{compoundness}.

There is another law which is important from a modeling perspective, namely `the marginal probability law'. This law refers to the possibility to attribute individual probabilities to the outcomes for each of the subentities, independently of the other subentity. Violation of the marginal law involves a `deeper' contextuality, meaning that in addition to the outcome of one subentity depending on the state of the other subentity, the probability of an outcome of this subentity depends on the state of the other subentity. The situation of a compound entity, with joint measurements and outcomes interpretable for the subentities, indeed allows to calculate separate probabilities for each outcome of one subentity. In accordance with the marginal law, such a calculation for one of the subentities must produce values independently of what happens with the other subentity. 

This law is violated in our lizard situation. Indeed, consider any color $a$ for one of the lizards, and call $p(a)_1$ the probability that this morph of color $a$ `wins'. We can then calculate the probability in different ways, depending on what the other morph is. Indeed, suppose that $b$ is a color different from $a$, then we have
\begin{eqnarray} 
&&p(a)_1=p(aa)_{11}+p(aa)_{12}=+1 \\
&&p(a)_1=p(ab)_{11}+p(ab)_{12}=p(ab)_{12}
\end{eqnarray}
And $p(ab)_{12}$ is different for different combinations of $a$ and $b$, and is only equal to 1 for some combinations in the ideal game. More specifically, for the solution that we calculated from the data in \cite{BleaySinervo2007}, and that we presented in the foregoing section in (\ref{solutionoy}), (\ref{solutionob}) and (\ref{solutionyb}), we have
\begin{eqnarray} \label{marginal}
&&p(y)_1=p(yo)_{11}+p(yo)_{12}=0.88 \\ \label{marginalagain}
&&p(y)_1=p(yb)_{11}+p(yb)_{12}=0.18 \\ \label{marginalagainn}
&&p(y)_1=p(yy)_{11}+p(yy)_{12}=1 
\end{eqnarray} 
which gives three different values for the probability that `a yellow morph wins', each value depending on what color the other morph is. Given the RPS dynamics, we can understand that when the other morph is orange this probability is bigger than when it is blue.

It can be shown that the violation of the marginal law -- in a way rather similar to the violation of Bell's Inequality -- also induces the presence of entanglement. However, as we will analyze in Section \ref{compoundness}, the entanglement caused by the violation of the marginal law is structurally such that it cannot be modeled in the state alone and also appears on the level of the measurements. 

In the next section, we start elaborating an explicit Hilbert space model providing our analysis with the necessary technical detail to make explicit all elements we have introduced so far.
 
\section{The construction of a Hilbert space model \label{hilbertspacemodel}}
In this section we construct an explicit Hilbert space model for the probabilities that we propose as a solution in (\ref{solutionob}), (\ref{solutionoy}) and (\ref{solutionyb}) for the data in \cite{BleaySinervo2007}. Parallel to its concrete construction, we will explain the mathematics of quantum modeling in Hilbert space, and the mathematics of Hilbert space itself. By proceeding in this parallel way, we can likewise put forward the essential elements of the type of model building we are engaging in here. It will also enable us to point out in detail, in Section \ref{kolmogorov}, the problem that is encountered when a classical Kolmogorovian model is attempted for this situation.

\bigskip
\noindent
{\bf 1.} When quantum theory is applied for modeling purposes, the entity to be modeled is associated with a complex Hilbert space ${\cal H}$.

\bigskip
\noindent
Before we explain in detail the way in which this association is made, let us specify what a complex Hilbert space ${\cal H}$ is. It is a vector space over the field ${\mathbb C}$ of complex numbers, equipped with an inner product $\langle \cdot | \cdot \rangle$ mapping two vectors $\langle u|$ and $|v\rangle$ to a complex number $\langle u|v\rangle$. We denote vectors by using the bra-ket notation introduced by Paul Adrien Dirac, one of the founding fathers of quantum theory \citep{Dirac1958}. Vectors can be kets, denoted by $|u\rangle $, $|v\rangle$, or bras, denoted by $\langle u|$, $\langle v|$. The inner product between the ket vectors $|u\rangle$ and $|v\rangle$, or the bra-vectors $\langle u|$ and $\langle v|$, is realized by juxtaposing the bra vector $\langle u|$ and the ket vector $|v\rangle$, and $\langle u|v\rangle$ is also called a bra-ket, and it satisfies the following properties: (i) $\langle u |  u \rangle \ge 0$; (ii) $\langle u |  v \rangle=\langle v |  u \rangle^{*}$, where $\langle v |  u \rangle^{*}$ is the complex conjugate of $\langle u |  v \rangle$; (iii) $\langle u |(z|v\rangle+t|w\rangle)=z\langle u |  v \rangle+t \langle u |  w \rangle $, for $z, t \in {\mathbb C}$,
where the sum vector $z|u\rangle+t|w\rangle$ is called a `superposition' of vectors $|u\rangle$ and $|w\rangle$ in the quantum jargon. From (ii) and (iii) follows that the bra-ket is linear in the ket and anti-linear in the bra, i.e. $(z\langle u|+t\langle v|)|w\rangle=z^{*}\langle u | w\rangle+t^{*}\langle v|w \rangle$.

For those not acquainted at all with the structure of a complex Hilbert space, but knowledgeable about vectors spaces in general over real numbers, we mention that a complex Hilbert space is exactly the same as a real vector space, except that real numbers are exchanged by complex numbers. The bra-ket replaces what is called the inner product of two vectors in a real vector space. Calculating in a complex Hilbert space is the same, except that the calculation rules or complex numbers need to be applied whenever numbers are multiplied. Let us add some more aspects of complex numbers, and the vectors of Hilbert space, that are needed for modeling.

We recall that the absolute value of a complex number is defined as the square root of the product of this complex number times its complex conjugate. In a formula, $|z|=\sqrt{z^{*}z}$.
A complex number $z$ can either be decomposed into its Cartesian form $z=x+iy$, or into its goniometric form $z=|z|e^{i\theta}=|z|(\cos\theta+i\sin\theta)$. Hence we have $|\langle u| v\rangle|=\sqrt{\langle u|v\rangle\langle v|u\rangle}$. We define the `length' of a ket (bra) vector $|u\rangle$ ($\langle u|$) as $|| |u\rangle ||=||\langle u |||=\sqrt{\langle u |u\rangle}$. A vector of unit length is called a `unit vector'. We say that the ket vectors $|u\rangle$ and $|v\rangle$ are `orthogonal' and write $|u\rangle \perp |v\rangle$ if $\langle u|v\rangle=0$. This introduces the necessary mathematics to describe the first modeling rule of quantum theory. 

\bigskip
\noindent
{\bf 2.} First modeling rule: The different situations that the modeled entity can be encountered in, which are called `states' by physicists, are represented by the ket vectors $|u\rangle$ of unit length, i.e. $\langle u|u\rangle=1$, hence unit vectors, of the complex Hilbert space ${\cal H}$ associated with the entity.

\bigskip
\noindent
Representing the situations, or states, of the considered entity is one fourth of the Hilbert space quantum representation procedures. The other parts consist of the rules and prescriptions to represent measurements, probabilities, and the composition of entities. We will explain these in the following, and also apply them parallel to our lizard situation in constructing a concrete quantum model for it. Hence, now we first need to introduce some additional mathematics of Hilbert space to explain the way in which measurements are represented.

Measurements are essentially represented by what are called `self-adjoint operators' on the Hilbert space ${\cal H}$. For the purpose of the Hilbert space representation we built for the lizard situation, we do not need to explain all mathematical aspects of this representation of measurements by self-adjoint operators. The situation we encounter is simpler than the general one, because we will work in a finite dimensional Hilbert space, while general quantum theory is made to cope with infinite dimensions. Each self-adjoint operator in a finite dimensional Hilbert space is uniquely determined by a set of vectors, which are called the `eigenvectors' of this self-adjoint operator. Each eigenvector corresponds to one of the outcomes, and the number of eigenvectors is also the number of dimensions of the Hilbert space. Each of such eigenvectors is orthogonal to each other one, and if we also decide to choose them of length equal to 1 -- which we always can -- then such a set of eigenvectors forms an orthonormal basis of the Hilbert space. With `basis', we mean what is ordinarily meant for an arbitrary vector space, i.e. a set of vectors such that each vector of the space can be written as a linear independent combination of these vectors. Such a combination is called a `superposition' in the quantum jargon. It is the existence of superposed states which is at the origin of the `interference effects' observed with quantum particles. We will construct the measurements by directly identifying, for each measurement, its orthonormal set of eigenvectors.

\bigskip
\noindent
{\bf 3.} Second modeling rule: A measurement is represented by a self-adjoint operator $H$ on ${\cal H}$, and, for a finite dimensional Hilbert space, this operator $H$ is determined by its set of eigenvectors, $\{|h_1\rangle, \ldots, |h_n\rangle\}$ which we can choose as an orthonormal basis of ${\cal H}$. Each eigenvector corresponds to an outcome of the measurement, and the dimension of the Hilbert space is determined by the number of eigenvectors of a typical measurement.

\bigskip
\noindent
Let us apply this second modeling rule to our lizards situation, and introduce the necessary notations, operators and vectors. We denote a measurement where a morph of color $a$ competes with a morph of color $b$, by means of the operator $H(ab)$. Such a measurements $H(ab)$ contains always four possible outcomes, i.e. `win, win', `win, lose', `lose, win' and `lose, lose'. Since each outcome requires a corresponding different eigenstate, we will have to determine for each measurement a basis of four orthonormal vectors, so that the Hilbert space we will use for our modeling is the four-dimensional complex Hilbert space. Let us introduce the notation for the eigenvectors.
For a measurement $H(ab)$ we introduce the orthonormal basis of eigenvectors
\begin{eqnarray}
&&|ab_{11}\rangle \quad |ab_{12}\rangle \quad |ab_{21}\rangle \quad |ab_{22}\rangle \\
&&\langle ab_{ij}|ab_{kl}\rangle=0 \quad {\rm for} \quad ij\not=kl \\
&&\langle ab_{ij}|ab_{ij}\rangle=1
\end{eqnarray}

\bigskip
\noindent
{\bf 4.} Next we have to introduce the modeling rule that introduces the probabilities. This can be achieved fairly simply using all the machinery previously introduced. The probabilities are determined as follows. Suppose the considered situation is represented by the unit vector $|u\rangle$ of the Hilbert space ${\cal H}$. For a measurement H, determined by its orthonormal basis of eigenvectors $\{|h_1\rangle, \ldots, |h_n\rangle\}$, the probability for an outcome corresponding to eigenvector $|h_m\rangle$ to occur, is given by $|\langle u|h_m\rangle|^2$, which is the square of the absolute value of the bra-ket between the state and the eigenvector.

\bigskip
\noindent
Let us apply this probability modeling rule to our lizard situation. There are nine different measurements to consider, namely $H(oo)$, $H(oy)$, $H(ob)$, $H(yo)$, $H(yy)$, $H(yb)$, $H(bo)$, $H(by)$ and $H(bb)$. For each of the nine measurements we have to construct a set of four orthonormal eigenvectors that satisfy the probability laws. From (\ref{drawprobability}) follows
\begin{eqnarray}
&&|\langle u|aa_{11}\rangle|^2=p(aa)_{11}=1 \label{aa_{11}} \\
&&|\langle u|aa_{12}\rangle|^2=p(aa)_{12}=0 \label{aa_{12}} \\
&&|\langle u|aa_{21}\rangle|^2=p(aa)_{21}=0 \label{aa_{21}} \\
&&|\langle u|aa_{22}\rangle|^2=p(aa)_{22}=0 \label{aa_{22}}
\end{eqnarray}
which gives us all the probability laws to be satisfied for the three measurements $H(oo)$, $H(yy)$ and $H(bb)$. Indeed, for equal-color confrontations, we have probability equal to 1 that a `win, win' outcome results. For the other color combinations, we use the symmetry and anti-symmetry conditions, and the probabilities that we have calculated in Section \ref{outcomeprobabilities}. Hence, from (\ref{antisymmetry}), (\ref{symmetry}), (\ref{solutionoy}), (\ref{solutionob}) and (\ref{solutionyb}) follows that, for measurement $H(oy)$, we have
\begin{eqnarray}
|\langle u|oy_{11}\rangle|^2=p(oy)_{11}=0 \label{oy_{11}} \\
|\langle u|oy_{12}\rangle|^2=p(oy)_{12}=0.12 \label{oy_{12}} \\
|\langle u|oy_{21}\rangle|^2=p(oy)_{21}=0.88 \label{oy_{21}} \\
|\langle u|oy_{22}\rangle|^2=p(oy)_{22}=0 \label{oy_{22}}
\end{eqnarray}
and, due to (\ref{symmetry}), we have for $H(yo)$
\begin{eqnarray}
|\langle u|yo_{11}\rangle|^2=p(yo)_{11}=0 \label{yo_{11}}\\
|\langle u|yo_{12}\rangle|^2=p(yo)_{12}=0.88 \label{yo_{12}}\\
|\langle u|yo_{21}\rangle|^2=p(yo)_{21}=0.12 \label{yo_{21}}\\
|\langle u|yo_{22}\rangle|^2=p(yo)_{22}=0 \label{yo_{22}}
\end{eqnarray}
For measurement $H(yb)$ we have
\begin{eqnarray}
|\langle u|yb_{11}\rangle|^2=p(yb)_{11}=0 \label{yb_{11}} \\
|\langle u|yb_{12}\rangle|^2=p(yb)_{12}=0.18 \label{yb_{12}} \\
|\langle u|yb_{21}\rangle|^2=p(yb)_{21}=0.82 \label{yb_{21}} \\
|\langle u|yb_{22}\rangle|^2=p(yb)_{22}=0 \label{yb_{22}}
\end{eqnarray}
and, due to (\ref{symmetry}), we have for $H(by)$
\begin{eqnarray}
|\langle u|by_{11}\rangle|^2=p(by)_{11}=0 \label{by_{11}} \\
|\langle u|by_{12}\rangle|^2=p(by)_{12}=0.82 \label{by_{12}} \\
|\langle u|by_{21}\rangle|^2=p(by)_{21}=0.18 \label{by_{21}} \\
|\langle u|by_{22}\rangle|^2=p(by)_{22}=0 \label{by_{22}}
\end{eqnarray}
And, for measurement $H(bo)$, we have
\begin{eqnarray}
|\langle u|bo_{11}\rangle|^2=p(bo)_{11}=0 \label{bo_{11}} \\
|\langle u|bo_{12}\rangle|^2=p(bo)_{12}=0.28 \label{bo_{12}} \\
|\langle u|bo_{21}\rangle|^2=p(bo)_{21}=0.72 \label{bo_{21}} \\
|\langle u|bo_{22}\rangle|^2=p(bo)_{22}=0 \label{bo_{22}}
\end{eqnarray}
and, due to (\ref{symmetry}), we have for $H(ob)$
\begin{eqnarray}
|\langle u|ob_{11}\rangle|^2=p(ob)_{11}=0 \label{ob_{11}} \\
|\langle u|ob_{12}\rangle|^2=p(ob)_{12}=0.72 \label{ob_{12}} \\
|\langle u|ob_{21}\rangle|^2=p(ob)_{21}=0.28 \label{ob_{21}} \\
|\langle u|ob_{22}\rangle|^2=p(ob)_{22}=0 \label{ob_{22}}
\end{eqnarray}
Equations (\ref{yo_{11}}) to (\ref{bo_{22}}) are all we need to construct the Hilbert space model.

We construct an explicit complex Hilbert space model by making use of the canonical complex Hilbert space ${\mathbb C}^4$, namely the set of all $4-$tuples of complex numbers, equipped with an addition and multiplication by a complex number, and an inner product defined as follows.
\begin{eqnarray} \label{C4}
&&{\mathbb C}^4=\{(z_1,z_2,z_3,z_4)\vert z_1,z_2,z_3,z_4 \in {\mathbb C} \} \\ \label{C4sum}
&&(z_1,z_2,z_3,z_4)+(z'_1,z'_2,z'_3,z'_4)=(z_1+z'_1,z_2+z'_2,z_3+z'_3,z_4+z'_4) \\ \label{C4product}
&&\lambda(z_1,z_2,z_3,z_4)=(\lambda z_1,\lambda z_2,\lambda z_3,\lambda z_4) \quad {\rm for} \quad \lambda \in {\mathbb C} \\ \label{C4braket}
&&\langle (z_1,z_2,z_3,z_4)|(z'_1,z'_2,z'_3,z'_4)\rangle=z^*_1z'_1+z^*_2z'_2+z^*_3z'_3+z^*_4z'_4
\end{eqnarray}
We start by choosing the unit ket vector $|u\rangle$ representing the situation of the lizards without any measurement being involved, by the first canonical base vector of ${\mathbb C}^4$. 
\begin{equation} \label{choiceofstate}
|u\rangle=(1,0,0,0)
\end{equation}
First we determine the unit vectors representing the eigenstates for the draw measurements $H(oo)$, $H(yy)$ and $H(bb)$, hence satisfying (\ref{aa_{11}}), (\ref{aa_{12}}), (\ref{aa_{21}}) and (\ref{aa_{22}}). We also make sure that each measurement is determined by a different set of eigenvectors. There is no unique solution, hence the one we present here is one of the possible solutions.
\begin{eqnarray}
&&|oo_{11}\rangle=(e^{-i\phi{oo}},0,0,0), |oo_{12}\rangle=(0,{1 \over \sqrt{2}},{1 \over \sqrt{2}},0), \nonumber \\
&&|oo_{21}\rangle=(0,{1 \over \sqrt{2}},-{1 \over \sqrt{2}},0), |oo_{22}\rangle=(0,0,0,1) \\
&&|yy_{11}\rangle=(e^{-i\phi{yy}},0,0,0), |yy_{12}\rangle=(0,{1 \over \sqrt{2}},0,{1 \over \sqrt{2}}), \nonumber \\
&&|yy_{21}\rangle=(0,{1 \over \sqrt{2}},0,-{1 \over \sqrt{2}}), |yy_{22}\rangle=(0,0,1,0) \\
&&|bb_{11}\rangle=(e^{-i\phi{bb}},0,0,0), |bb_{12}\rangle=(0,0,{1 \over \sqrt{2}},{1 \over \sqrt{2}}), \nonumber \\
&&|bb_{21}\rangle=(0,0,{1 \over \sqrt{2}},-{1 \over \sqrt{2}}), |bb_{22}\rangle=(0,1,0,0)
\end{eqnarray}
We can easily verify that the above choices for the eigenvectors constitute an orthonormal basis for each of the measurements and satisfy (\ref{aa_{11}}), (\ref{aa_{12}}), (\ref{aa_{21}}) and (\ref{aa_{22}}). Let us check some of them to see how this works. Applying (\ref{C4braket}), we have
\begin{eqnarray}
\langle oo_{11}|oo_{21}\rangle&=&\langle (e^{-i\phi{oo}},0,0,0)|(0,{1 \over \sqrt{2}},-{1 \over \sqrt{2}},0)\rangle \nonumber \\
&=&(e^{-i\phi{oo}})^*(0)+(0)({1 \over \sqrt{2}})+(0)(-{1 \over \sqrt{2}})=0 \\
|\langle u|oo_{11}\rangle|^2&=&|\langle (1,0,0,0)|(e^{-i\phi{oo}},0,0,0)\rangle|^2 \nonumber \\
&=&|e^{-i\phi{oo}}|^2=1
\end{eqnarray}
The first shows that these two vectors are orthogonal. To prove orthonormality, we need to check also the orthogonality with the others and show that their lengths equal 1, but these calculations are analogous. The second shows that (\ref{aa_{11}}) is satisfied for $a$ being orange. All the others, i.e. (\ref{aa_{12}}), (\ref{aa_{21}}) and (\ref{aa_{22}}), and for all other colors, are proven analogously.

The eigenvectors for the remaining six measurements $H(oy)$, $H(ob)$, $H(yo)$, $H(yb)$, $H(bo)$ and $H(by)$ are more difficult to determine, because their probabilities do not follow just from the properties of the situation itself, but contain also the traces of the real-world data measured in \cite{BleaySinervo2007}. Let us make the construction for $H(yo)$. Some simplification still follows from the overall structure of the situation. For example, because of (\ref{yo_{11}}) and (\ref{yo_{22}}), we can take, without loss of generality
\begin{eqnarray}
|yo_{11}\rangle=(0,{1 \over \sqrt{2}},{1 \over \sqrt{2}},0)\\
|yo_{22}\rangle=(0,{1 \over \sqrt{2}},-{1 \over \sqrt{2}},0)
\end{eqnarray}
and look for $|yo_{12}\rangle$ and $|yo_{21}\rangle$ for a solution, in the form
\begin{eqnarray}
|yo_{12}\rangle=(a,0,0,be^{i\beta})\\
|yo_{21}\rangle=(c,0,0,de^{i\delta})
\end{eqnarray}
with $a^2+b^2=c^2+d^2=1$, which is needed for the vector to have a length equal to 1, and such that (\ref{yo_{12}}) and (\ref{yo_{21}}) are satisfied, which means that
\begin{eqnarray}
0.88=|\langle u|yo_{12}\rangle|^2=\langle (1,0,0,0)|(a,0,0,be^{i\beta})\rangle=a^2 \\
0.12=|\langle u|yo_{21}\rangle|^2=\langle (1,0,0,0)|(c,0,0,de^{i\delta})\rangle=c^2
\end{eqnarray}
Hence, a solution is given by $a=\sqrt{0.88}=0.94$, and $c=\sqrt{0.12}=0.34$. Then $b=\sqrt{1-a^2}=\sqrt{1-0.88}=\sqrt{0.12}=0.34$, and $d=\sqrt{1-c^2}=\sqrt{1-0.12}=\sqrt{0.88}=0.94$. This gives us
\begin{eqnarray} \label{withphase01}
|yo_{12}\rangle=(0.94,0,0,0.34 \cdot e^{i\beta}) \\ \label{withphase02}
|yo_{21}\rangle=(0.34,0,0,0.94 \cdot e^{i\delta})
\end{eqnarray}
The phases $\beta$ and $\delta$ can now be chosen to make these vectors orthogonal. 
We choose $\beta=0$ and $\delta=\pi$, and hence $e^{i\beta}=+1$ and $e^{i\delta}=-1$. This gives us the following solution for the orthonormal set of eigenvectors of the measurement $H(yo)$.
\begin{eqnarray}
|yo_{11}\rangle=(0,{1 \over \sqrt{2}},{1 \over \sqrt{2}},0)\\
|yo_{12}\rangle=(0.94,0,0,0.34) \label{withoutphase01} \\
|yo_{21}\rangle=(0.34,0,0,-0.94) \label{withoutphase02} \\
|yo_{22}\rangle=(0,{1 \over \sqrt{2}},-{1 \over \sqrt{2}},0)
\end{eqnarray}
In a very analogous way, we construct a solution for the measurement $H(by)$, but starting from
\begin{eqnarray}
|by_{11}\rangle=(0,{1 \over \sqrt{2}},0,{1 \over \sqrt{2}}) \\
|by_{22}\rangle=(0,{1 \over \sqrt{2}},0,-{1 \over \sqrt{2}})
\end{eqnarray} 
and hence we get
\begin{eqnarray}
|by_{12}\rangle=(0.91,0,0.42,0) \\
|by_{21}\rangle=(0.42,0,-0.91,0)
\end{eqnarray}
where $0.91=\sqrt{0.82}$ and $0.42=\sqrt{0.18}$.
Hence the orthornormal set of eigenvectors of $H(by)$ can be taken to be
\begin{eqnarray}
|by_{11}\rangle=(0,{1 \over \sqrt{2}},0,{1 \over \sqrt{2}}) \\
|by_{12}\rangle=(0.91,0,0.42,0) \\
|by_{21}\rangle=(0.42,0,-0.91,0) \\
|by_{22}\rangle=(0,{1 \over \sqrt{2}},0,-{1 \over \sqrt{2}})
\end{eqnarray} 
For the measurement $H(ob)$ we analogously construct the following orthonormal set of eigenvectors
\begin{eqnarray}
|ob_{11}\rangle=(0,0,{1 \over \sqrt{2}},{1 \over \sqrt{2}}) \\
|ob_{12}\rangle=(0.85,0.53,0,0) \\
|ob_{21}\rangle=(0.53,-0.85,0,0) \\
|ob_{22}\rangle=(0,0,{1 \over \sqrt{2}},-{1 \over \sqrt{2}})
\end{eqnarray}
where $0.85=\sqrt{0.72}$ and $0.53=\sqrt{0.28}$.

Let us now also construct orthonormal sets of eigenvectors of the remaining measurements $H(oy)$, $H(yb)$ and $H(bo)$. Let us first consider $H(oy)$. Remark that, because of $p(yo)_{12}=p(oy)_{21}=0.88$, and $p(yo)_{21}=p(oy)_{12}=0.12$, in the construction procedure we adapted for the three foregoing measurements, we might consider the two measurements to be equivalent. This is in fact even the case for the situations that we have been considering. Let us show, however, that Hilbert space allows us to make a construction such that both measurements are described by a different orthonormal basis. If in the future more refined measurements on the situation are made, it may prove necessary to distinguish both measurements.
It is also a way to show how the phases and hence the complex nature of the Hilbert space play a role in this possibility of distinguishing between $H(oy)$ and $H(yo)$, although there is no difference with regard to the probabilities that have been measured. Indeed, instead of choosing $\beta=0$ and $\delta=\pi$, in the step we made from (\ref{withphase01}) and (\ref{withphase02}), to (\ref{withoutphase01}) and (\ref{withoutphase02}), we can choose $\beta={\pi \over 2}$, and $\delta={3\pi \over 2}$. It is a property of the two-dimensional complex Hilbert space, which is the subspace we are working in with this choice of phase, that there are these different ways for vectors to be orthogonal, namely whenever the difference in angle between the phases equals $\pi$. Let us also recall that $e^{i{\pi \over 2}}=i$, and $e^{i{3\pi \over 2}}=-i$. This choice gives us the following orthonormal basis of eigenvectors for $H(oy)$.
\begin{eqnarray}
|oy_{11}\rangle=(0,{1 \over \sqrt{2}},{i \over \sqrt{2}},0)\\
|oy_{12}\rangle=(0.34,0,0,-i\cdot 0.94) \label{withoutphase001} \\
|oy_{21}\rangle=(0.94,0,0,i \cdot 0.34) \label{withoutphase002} \\
|oy_{22}\rangle=(0,{1 \over \sqrt{2}},-{i \over \sqrt{2}},0)
\end{eqnarray}
and in an analogous way we construct the orthonormal basis of eigenvectors for the two remaining measurements $H(yb)$ and $H(bo)$. They are
\begin{eqnarray}
|yb_{11}\rangle=(0,{1 \over \sqrt{2}},0,{i \over \sqrt{2}}) \\
|yb_{12}\rangle=(0.42,0,-i \cdot 0.91,0) \\
|yb_{21}\rangle=(0.91,0,i \cdot 0.42,0) \\
|yb_{22}\rangle=(0,{1 \over \sqrt{2}},0,-{i \over \sqrt{2}})
\end{eqnarray} 
\begin{eqnarray}
|bo_{11}\rangle=(0,0,{1 \over \sqrt{2}},{i \over \sqrt{2}}) \\
|bo_{12}\rangle=(0.53,-i \cdot 0.85,0,0) \\
|bo_{21}\rangle=(0.85,i \cdot 0.53,0,0) \\
|bo_{22}\rangle=(0,0,{1 \over \sqrt{2}},-{i \over \sqrt{2}})
\end{eqnarray}
This completes the construction of the orthonormal basis for all measurements.

In our explanation of the use of the quantum formalism for modeling purposes, we have introduced a measurement as characterized by an orthonormal basis. We did mention that the common way to represent a measurement is by a self-adjoint operator, and that is how one will find it described in quantum theory textbooks. The step from the orthonormal basis to a self-adjoint operator is straightforward in the case of a finite dimensional Hilbert space. Indeed, such a self-adjoint operator can then be represented as a square Hermitian matrix $H(ab)_{ij}$, with size equal to the dimension of the Hilbert space used. Hermitian means that the diagonal elements are real, and the off-diagonal elements are complex conjugates, i.e. $H(ab)_{ij}=H(ab)_{ji}^*$. The vectors representing a measurement as an orthonormal basis are the eigenvectors of this matrix. When the matrix is diagonalised, and hence contains only real numbers on its diagonal, and zero's for all its other elements, these real numbers are the eigenvalues. These eigenvalues are numbers given to identify the different outcomes, hence if we want to distinguish the four different outcomes in each one of our joint experiments, we need to choose for each self-adjoint operator four different real numbers $\lambda_1, \lambda_2, \lambda_3, \lambda_4$, each of the numbers characterising one of the four outcomes. 

We will calculate two of such self-adjoint operators for the lizard situation, because we want to show explicitly that they do not commute. Non-commuting self-adjoint operators are indeed the hallmark of non-classicality of the probability model within the quantum formalism. We calculate the matrix corresponding to a set of eigenvectors, and eigenvalues, using standard techniques from linear algebra, as follows

\footnotesize
\begin{eqnarray}
H(by)&=&
\lambda_1\cdot \left( \begin{array}{c}
0	\\			
{1 \over \sqrt{2}}  \\	
{1 \over \sqrt{2}}  \\	
0  \end{array} \right)  \cdot \left( \begin{array}{cccc}
0	& {1 \over \sqrt{2}} & {1 \over \sqrt{2}} & 0
\end{array} \right)+\lambda_2\cdot \left( \begin{array}{c}
0.91	\\			
0  \\	
0  \\	
0.42  \end{array} \right) \cdot \left( \begin{array}{cccc}
0.91	& 0 & 0 & 0.42
\end{array} \right) \nonumber \\
&&+\lambda_3\cdot  \left( \begin{array}{c}
0.42	\\			
0  \\	
0   \\	
-0.91  \end{array} \right) \cdot \left( \begin{array}{cccc}
0.42	& 0 & 0 & -0.91
\end{array} \right)+\lambda_4\cdot \left( \begin{array}{c}
0	\\			
{1 \over \sqrt{2}}  \\	
-{1 \over \sqrt{2}}  \\	
0  \end{array} \right) \cdot \left( \begin{array}{cccc}
0	& {1 \over \sqrt{2}} & -{1 \over \sqrt{2}} & 0
\end{array} \right) \nonumber 
\end{eqnarray}
\begin{eqnarray} 
&=&\left( \begin{array}{cccc}
0	& 0 & 0 & 0	\\			
0 & {1 \over 2}\lambda_1 & {1 \over 2}\lambda_1 & 0 \\	
0 & {1 \over 2}\lambda_1 & {1 \over 2}\lambda_1	& 0 \\	
0 &	0 &	0 & 0 \end{array} \right) + \left( \begin{array}{cccc}
0.82 \lambda_2	& 0 & 0 & 0.38	 \lambda_2\\			
0 & 0 & 0 & 0 \\	
0 & 0 & 0	& 0 \\	
0.38 \lambda_2 &	0 &	0 & 0.18 \lambda_2 \end{array} \right) \nonumber \\
&&+\left( \begin{array}{cccc}
0.18 \lambda_3	& 0 & 0 & -0.38 \lambda_3	\\			
0 & 0 & 0 & 0 \\	
0 & 0 & 0	& 0 \\	
-0.38 \lambda_3 &	0 &	0 & 0.82 \lambda_3 \end{array} \right)+\left( \begin{array}{cccc}
0	& 0 & 0 & 0	\\			
0 & {1 \over 2}\lambda_4 & -{1 \over 2}\lambda_4 & 0 \\	
0 & -{1 \over 2}\lambda_4 & {1 \over 2}\lambda_4	& 0 \\	
0 &	0 &	0 & 0 \end{array} \right) \nonumber \\
&=&\left( \begin{array}{cccc}
0.82 \lambda_2+0.18 \lambda_3	& 0 & 0 & 0.38 (\lambda_2-\lambda_3)	\\			
0 & {1 \over 2}(\lambda_1+\lambda_4) & {1 \over 2}(\lambda_1-\lambda_4) & 0 \\	
0 & {1 \over 2}(\lambda_1-\lambda_4) & {1 \over 2}(\lambda_1+\lambda_4)	& 0 \\	
0.38 (\lambda_2-\lambda_3) &	0 &	0 & 0.18 \lambda_2+0.82 \lambda_3 \end{array} \right)
\end{eqnarray}
\normalsize
where $\lambda_1, \lambda_2, \lambda_3$ and $\lambda_4$ are four different real numbers that identify respectively the outcomes `win, win', `win, lose', `lose, win' and `lose, lose' for a competition of two morphs of color `blue' and `yellow'. 
In an analogous way we calculate, for example, $H(bo)$

\footnotesize
\begin{eqnarray}
H(bo)&=&
\mu_1\cdot \left( \begin{array}{c}
0	\\			
{1 \over \sqrt{2}}  \\	
0  \\	
{1 \over \sqrt{2}}  \end{array} \right)  \cdot \left( \begin{array}{cccc}
0	& {1 \over \sqrt{2}} & 0 & {1 \over \sqrt{2}}
\end{array} \right)+\mu_2\cdot \left( \begin{array}{c}
0.85	\\			
0  \\	
0.53  \\	
0  \end{array} \right) \cdot \left( \begin{array}{cccc}
0.85	& 0 & 0.53 & 0
\end{array} \right) \nonumber \\
&&+\mu_3\cdot  \left( \begin{array}{c}
0.53	\\			
0  \\	
-0.85   \\	
0  \end{array} \right) \cdot \left( \begin{array}{cccc}
0.53	& 0 & -0.85 & 0
\end{array} \right)+\mu_4\cdot \left( \begin{array}{c}
0	\\			
{1 \over \sqrt{2}}  \\	
0  \\	
-{1 \over \sqrt{2}}  \end{array} \right) \cdot \left( \begin{array}{cccc}
0	& {1 \over \sqrt{2}} & 0 & -{1 \over \sqrt{2}}
\end{array} \right) \nonumber \\
&=&\left( \begin{array}{cccc}
0	& 0 & 0 & 0	\\			
0 & {1 \over 2}\mu_1 & 0 & {1 \over 2}\mu_1 \\	
0 & 0 & 0	& 0 \\	
0 &	{1 \over 2}\mu_1 &	0 & {1 \over 2}\mu_1 \end{array} \right) + \left( \begin{array}{cccc}
0.72\mu_2	& 0 & 0.45\mu_2 & 0	\\			
0 & 0 & 0 & 0 \\	
0.45\mu_2 & 0 & 0.28\mu_2	& 0 \\	
0 &	0 &	0 & 0 \end{array} \right) \nonumber \\
&&+\left( \begin{array}{cccc}
0.28\mu_3	& 0 & -0.45\mu_3 & 0	\\			
0 & 0 & 0 & 0 \\	
-0.45\mu_3 & 0 & 0.72\mu_3	& 0 \\	
0 &	0 &	0 & 0 \end{array} \right)+\left( \begin{array}{cccc}
0	& 0 & 0 & 0	\\			
0 & {1 \over 2}\mu_4 & 0 & -{1 \over 2}\mu_4 \\	
0 & 0 &  0	& 0 \\	
0 &	-{1 \over 2}\mu_4 &	0 & {1 \over 2}\mu_4 \end{array} \right) \nonumber 
\end{eqnarray}
\begin{eqnarray}
&=&\left( \begin{array}{cccc}
0.72\mu_2+0.28\mu_3	& 0 & 0.45(\mu_2-\mu_3) & 0	\\			
0 & {1 \over 2}(\mu_1+\mu_4) & 0 & {1 \over 2}(\mu_1-\mu_4) \\	
0.45(\mu_2-\mu_3) & 0 & 0.72\mu_2+0.28\mu_3	& 0 \\	
0 &	{1 \over 2}(\mu_1-\mu_4) &	0 & {1 \over 2}(\mu_1+\mu_4) \end{array} \right)
\end{eqnarray}
\normalsize
where this time $\mu_1, \mu_2, \mu_3$ and $\mu_4$ are four different real numbers, identifying respectively the outcome `win, win', `win, lose', `lose, win' and `lose, lose' of a competition of two morphs with colors `blue' and `orange'.
Let us show that $H(yo)$ and $H(by)$ do not commute. We have
\begin{eqnarray}
H(yo)H(by)_{ij}=\sum_kH(yo)_{ik}H(by)_{kj}
\end{eqnarray}
Consider for example the element in the first column and second row of the two products, i.e. the elements $H(yo)H(by)_{21}$ and $H(by)H(yo)_{21}$. We have
\begin{eqnarray}
H(yo)H(by)_{21}&=&H(yo)_{21}H(by)_{11}+H(yo)_{22}H(by)_{21}\nonumber \\
&+&H(yo)_{23}H(by)_{31}+H(yo)_{24}H(by)_{41} \nonumber \\ \label{product01}
&=&{1 \over 2}(\lambda_1-\lambda_4)\cdot0.45(\mu_2-\mu_3) \\
H(by)H(yo)_{21}&=&H(by)_{21}H(yo)_{11}+H(by)_{22}H(yo)_{21} \nonumber \\
&+&H(by)_{23}H(yo)_{31}+H(by)_{24}H(yo)_{41} \nonumber \\ \label{product02}
&=&{1 \over 2}(\mu_1-\mu_4)\cdot 0.38(\lambda_2-\lambda_3) 
\end{eqnarray}
Let us mention here that the commutation of two self-adjoint operators in the way they represent measurements in quantum theory is equivalent to the commutation of the projectors on their orthonormal base of eigenvectors, which shows that the specific value of the eigenvalues plays no role in it. Hence, it is sufficient to observe that (\ref{product01}) is different from (\ref{product02}) for specific values of $\lambda_1, \lambda_2, \lambda_3, \lambda_4$ and $\mu_1, \mu_2, \mu_3$ and $\mu_4$, to conclude about the non commutativity of the two self-adjoint operators representing the joint measurements for `blue' and `yellow' competing morphs, and `yellow' and `orange' competing morphs.  

\section{Quantum compoundness, subentities, submeasurements and entanglement\label{compoundness}}
We have constructed a four dimensional Hilbert space model for the compound entity consisting of the two interacting lizards. In this section we will analyze the way in which the two individual lizards and the measurements we have defined with respect to their winning and losing appear as subentities and submeasurements, and show that quantum entanglement is involved for both.

Contrary to classical theory, where compoundness appears in a way equivalent to how two subsets are `joined' by means of the `joining of subsets', and hence no `new is added', in quantum theory compoundness involves the emergence of new states and new measurements. These new states and measurements arise as a consequence of the mathematical structure of quantum theory. It is indeed the vector space structure of the set of states and the linear algebra structure of the set of measurements that generate these new states and measurements for a situation of compoundness. It is, in effect, this property of emergence which makes quantum-like structures better suited as compared to classical structures to model situations of compoundness in the natural world. Indeed, in a natural situation usually new states and new measurements arise when two entities are joined. Since in a situation modeled by classical structures compoundness is reduced to a simple union of the existing subentities and submeasurements, these subentities and submeasurements can easily be retrieved from the structure of the compound entity and measurements. In a situation of the natural world, and certainly so in a situation modeled by quantum theory, retrieving the subentities and submeasurements is more complicated, and involves complex aspects due to the effect of emergence. Hence, to analyze this situation of compoundness, we carefully employ the mathematical procedures of quantum theory designed for retrieving the subentities and measurements. As we will see, the two lizards, interacting following RPS-like internal dynamics, constitute an example of non-classical compoundness in a very significant way. The aim of the present section is to investigate this situation in detail.

We continue the approach we initiated in Section \ref{hilbertspacemodel}, and introduce the fourth quantum modeling rule, explaining how compound entities are analyzed in function of their constituting subentities, while in parallel we investigate the situation of the lizards with respect to compoundness. 

\bigskip
\noindent
{\bf 3.} Suppose that the compound entity $S$ is made up of two subentities $S_1$ and $S_2$, and that $S$, $S_1$ and $S_2$ are described by the complex Hilbert spaces ${\cal H}$, ${\cal H}_1$ and ${\cal H}_2$, respectively, following the standard quantum formalism, i.e. a modeling as explained in Section \ref{hilbertspacemodel}. Identifying the subentities and submeasurements consists in considering an isomorphism between the Hilbert space ${\cal H}$ and the tensor product ${\cal H}_1 \otimes {\cal H}_2$ of the two Hilbert spaces ${\cal H}_1$ and ${\cal H}_2$. The image of this isomorphism of states and measurements is interpreted in this tensor product. Entanglement in a state indicates the situation where the image of this state cannot be written as a product state in the tensor product.

The tensor product is in many ways the only possible structure to constitute the basis for the description of the compound entity of two subentities within the quantum formalism. It is the Hilbert space generated linearly by the product states, but it can also be proven to model the compound entity from an operational axiomatic point of view \citep{AertsDaubechies1978}. 

To model the compound entity of the two interacting lizards, we have explicitly introduced the four-dimensional complex Hilbert space in its canonical form ${\mathbb C}^4$. Each of the lizards is individually modeled as a subentity in a two-dimensional complex Hilbert space, and its canonical form is ${\mathbb C}^2$. The tensor product ${\mathbb C}^2 \otimes{\mathbb C}^2$ is a four-dimensional complex Hilbert space, which is to be used to identify the subentities and submeasurements of the entity of the interacting lizards. It is defined as follows
\begin{eqnarray}
{\mathbb C}^2 \otimes{\mathbb C}^2=\{\sum_{ij}\lambda_{ij}|v\rangle_i\otimes |w\rangle_j\ \vert\ |v\rangle_i, |w\rangle_j \in {\mathbb C}^2, \lambda_{ij} \in {\mathbb C}\}
\end{eqnarray} 
where the tensor product $\otimes$ is an operation with the usual properties of a product, i.e. for $|u\rangle, |v\rangle, |w\rangle, |t\rangle \in {\mathbb C}^2$ and $\lambda, \mu, \nu, \theta \in {\mathbb C}$, we have 
\begin{eqnarray}
&& (\lambda|u\rangle+\mu|v\rangle) \otimes (\nu|w\rangle+\theta|t\rangle) \nonumber \\
&&=\lambda\nu |v\rangle \otimes |w\rangle+\mu\nu|v\rangle \otimes |w\rangle+\lambda \theta |v\rangle \otimes |t\rangle+\mu\theta|v\rangle \otimes |t\rangle
\end{eqnarray}
The bra-ket is defined as follows
\begin{eqnarray}
(\langle u| \otimes \langle v|)(|w\rangle \otimes |t\rangle)=\langle u|w\rangle \langle v|t\rangle
\end{eqnarray}
It can be verified straightforwardly that if $\{|o_1\rangle,|o_2\rangle\}$ and $\{|y_1\rangle,|y_2\rangle\}$ are orthonormal bases of ${\mathbb C}^2$, then $\{|o_1\rangle\otimes |y_1\rangle, |o_1\rangle\otimes |y_2\rangle, |o_2\rangle\otimes |y_1\rangle,|o_2\rangle\otimes |y_2\rangle\}$ is an orthonormal basis of ${\mathbb C}^2\otimes {\mathbb C}^2$.
The tensor product is not commutative, i.e. in general we have
\begin{equation}
|v\rangle \otimes |w\rangle \not= |w\rangle \otimes |v\rangle
\end{equation}

We make explicit what entanglement is on the level of the states. Suppose we have constructed an orthonormal basis $\{|o_1\rangle\otimes |y_1\rangle, |o_1\rangle\otimes |y_2\rangle, |o_2\rangle\otimes |y_1\rangle,|o_2\rangle\otimes |y_2\rangle\}$, starting from two different orthonormal bases $\{|o_1\rangle,|o_2\rangle\}$ and $\{|y_1\rangle,|y_2\rangle\}$ of ${\mathbb C}^2$, and, for $\lambda_{12}$ and $\lambda_{21}$ both different from zero, consider the following vector of ${\mathbb C}^2\otimes {\mathbb C}^2$, 
\begin{eqnarray} \label{entangledvector}
|u\rangle=\lambda_{12}|o_1\rangle \otimes |y_2\rangle+\lambda_{21}|o_2\rangle \otimes |y_1\rangle
\end{eqnarray} 
then it is not possible to find vectors $|v\rangle, |w\rangle \in {\mathbb C}^2$, such that $|u\rangle=|v\rangle \otimes |w\rangle$ is a product of such two vectors. This can be shown right away, but let us try it out. Suppose that we are looking for two such vectors $|v\rangle, |w\rangle \in {\mathbb C}^2$, such that $|u\rangle=|v\rangle \otimes |w\rangle$. Then, since $\{|o_1\rangle,|o_2\rangle\}$ and $\{|y_1\rangle,|y_2\rangle\}$ are both bases of ${\mathbb C}^2$, we can write
\begin{eqnarray}
|v\rangle=\lambda_1|o_1\rangle+\lambda_2|o_2\rangle \\
|w\rangle=\mu_1|y_1\rangle+\mu_2|y_2\rangle
\end{eqnarray}
where, for $\lambda_1, \lambda_2$, and $\mu_1, \mu_2$, for each couple, at least one of them is different from zero. Then we have
\begin{eqnarray}
|v\rangle \otimes |w\rangle&=&(\lambda_1|o_1\rangle+\lambda_2|o_2\rangle) \otimes (\mu_1|y_1\rangle+\mu_2|y_2\rangle) \\
&=&\lambda_1\mu_1 |o_1\rangle \otimes |y_1\rangle + \lambda_1\mu_2 |o_1\rangle \otimes |y_2\rangle \nonumber \\
&&+ \lambda_2\mu_1 |o_2\rangle \otimes |y_1\rangle + \lambda_2\mu_2 |o_2\rangle \otimes |y_2\rangle
\end{eqnarray}
Since $\{|o_1\rangle\otimes |y_1\rangle, |o_1\rangle\otimes |y_2\rangle, |o_2\rangle\otimes |y_1\rangle,|o_2\rangle\otimes |y_2\rangle\}$ is a an orthonormal basis of ${\mathbb C}^2\otimes {\mathbb C}^2$, from $|u\rangle=|v\rangle \otimes |w\rangle$ then follows that
\begin{eqnarray}
\lambda_1\mu_1=0 \quad \lambda_1\mu_2=\lambda_{12} \quad \lambda_2\mu_1=\lambda_{21} \quad \lambda_2\mu_2=0
\end{eqnarray}
which are four equations impossible to be satisfied together for the four complex numbers $\lambda_1$, $\lambda_2$, $\mu_1$ and $\mu_2$. Indeed, from $\lambda_1\mu_1=0$ 
follows that $\lambda_1=0$ or $\mu_1=0$, and hence at least one of $\lambda_{12}$ or $\lambda_{21}$ is zero, which by construction of the vector $|u\rangle$ is not the case.

If the compound entity is in such a state $|u\rangle$, this means that it cannot be interpreted as the two subentities being in individual states each of them. Hence $|u\rangle$ is an example of a new emergent state of the compound entity.

Let us show that it is exactly states of this nature that appear in our modeling of the lizard interactions, when we introduce the tensor product procedure to identify the subentities and submeasurements of the compound lizard entity. Suppose that for the individual lizards, using the complex Hilbert space ${\mathbb C}^2$ to represent its states, we model the measurements that can result in an individual win or lose, by the orthonormal basis $\{|o_1\rangle, |o_2\rangle\}$, $\{|y_1\rangle, |y_2\rangle\}$, and $\{|b_1\rangle, |b_2\rangle\}$, for the different colors. Concretely, this means that in case one of the two lizards is in state $|v\rangle \in {\mathbb C}^2$, and the other one in state $|w\rangle \in {\mathbb C}^2$, then $o_1$ is the outcome of an orange morph that wins, and $o_2$ the outcome of an orange morph that loses. Similarly, $y_1$ is the outcome of a yellow morph that wins, and $y_2$ is the outcome of a yellow morph that loses, and $b_1$ is the outcome of a blue morph that wins, while $b_2$ is the outcome of a blue morph that loses. 

Taking into account our construction of the orthonormal basis for measurements on individual lizards, it follows that $|\langle v|o_1\rangle|^2$ ($|\langle w|o_1\rangle|^2$) is the probability for
the first (the second) individual orange morph to win, $|\langle v|
o_2\rangle|^2$ ($|\langle w|o_2\rangle|^2$) for the first (the second) individual orange morph to lose. Similarly, $|\langle u|y_1\rangle|^2$ is the probability for the first (the second) individual yellow morph to win, $|\langle v|y_2\rangle|^2$ ($|\langle w|y_2\rangle|^2$) for the first (the second) individual yellow morph to lose, and $|\langle v|b_1\rangle|^2$ ($|\langle w|b_1\rangle|^2$) for the first (the second) individual blue morph to win, while $|\langle v|b_2\rangle|^2$ ($|\langle w|b_2\rangle|^2$) for the first (the second) individual blue morph to lose. This means that we apply the quantum Hilbert space formalism, and all of its modeling rules, to individual morphs. 

We should point out that, since the physical animals which are the individual lizards, cannot change colors -- although there is some evidence that suggests that the polymorphism of the \emph{Uta stansburiana} might occasionally also be environmentally triggered --, and hence cannot change strategies, we do not model the physical animal itself. In our model, a morph's color is therefore considered to be a variable. One interpretation is to imagine a specific `competition situation' which consists in two morphs appearing in this situation competing for a female. The potential states of these two morphs in this situation are those described in the model, and color is included in this potential. The different possible competition situations as such are modeled by $|v\rangle$ for one of the morphs, and by $|w\rangle$ for the other morph, and by $|u\rangle$ for the compound of the two morphs.

We propose the following isomorphism with the tensor product space.
\begin{eqnarray}
&&I_{oy}: {\mathbb C}^4 \rightarrow {\mathbb C}^2 \otimes{\mathbb C}^2 \\
&&I_{oy}|oy_{11}\rangle=|o_1\rangle \otimes |y_1\rangle \quad I_{oy}|oy_{12}\rangle=|o_1\rangle \otimes |y_2\rangle \\
&&I_{oy}|oy_{21}\rangle=|o_2\rangle \otimes |y_1\rangle \quad I_{oy}|oy_{22}\rangle=|o_2\rangle \otimes |y_2\rangle
\end{eqnarray}
Let us calculate the image of the state $|u\rangle$, respecting, of course, our choice for the Hilbert space modeling regarding the competition compound entity situation in Section \ref{hilbertspacemodel}. We recall that we choose $|u\rangle=(1,0,0,0)$ as specified and explained in (\ref{choiceofstate}). To calculate the image of $|u\rangle$, let us write $|u\rangle$ as a linear combination in the orthonormal basis $\{|oy_{11}\rangle, |oy_{12}\rangle, |oy_{21}\rangle, |oy_{22}\rangle\}$ of ${\mathbb C}^4$, because the isomorphism $I_{oy}$ is defined by its actions on these vectors. We have
\begin{eqnarray}
|u\rangle&=&|oy_{11}\rangle \langle oy_{11}|u\rangle+|oy_{12}\rangle \langle oy_{12}|u\rangle \nonumber \\
&&+|oy_{21}\rangle \langle oy_{21}|u\rangle+|oy_{22}\rangle \langle oy_{22}|u\rangle \\
&=&0.34|oy_{12}\rangle +0.94|oy_{21}\rangle
\end{eqnarray}
where we used (\ref{withoutphase001}) and (\ref{withoutphase002}).
This means that we have 
\begin{eqnarray}
I_{oy}|u\rangle&=&I_{oy}(0.334|oy_{12}\rangle +0.994|oy_{21}\rangle) \\
&=&0.34 I_{oy}|oy_{12}\rangle +0.94 I_{oy}|oy_{21}\rangle \\
&=&0.34 |o_1\rangle \otimes |y_2\rangle +0.94 |o_2\rangle \otimes |y_1\rangle
\end{eqnarray}
This is exactly the type of vector which we introduced in (\ref{entangledvector}), with $\lambda_{12}=0.34$ and $\lambda_{21}=0.94$, and we proved it to be entangled.

It can be shown that the image of the self-adjoint operator $H_{oy}$ representing the orange yellow competition measurement is a product operator. This means that the entanglement of the orange yellow competition situation can be fully entered into the state. However, if we want to accomplish this for other color combination competitions, we will have to consider different isomorphisms, one specific one for each color combination. This is due to the marginal probability law being violated, like we have shown in (\ref{marginal}), (\ref{marginalagain}) and (\ref{marginalagainn}). 
We have analyzed this aspect of an entanglement situation in detail in \cite{asIQSA2012}.

\section{Non-Kolmogorovity in the RPS game\label{kolmogorov}}
We mentioned in Section \ref{intro} 
that it was the violation of Bell's inequalities for the RPS dynamics that caught our attention, and more specifically that such a violation indicates the presence of a non-Kolmogorovian structure for the considered probability model \citep{AccardiFedullo1982,Pitowsky1989}. In this section we analyze the concrete meaning of this violation by constructing a Kolmogorovian model and identifying where and why it fails. Again, of course, this analysis 
is from the specific viewpoint of considering the two individual lizard morphs as subentities of the compound lizard system.

Hence, suppose there exists a Kolmogorovian model for the situation of competing lizards considered as a compound situation of two lizards. 
This means that we have a sample space $\Omega$, and a probability measure $\mu$ on $\Omega$. The situation of one of the lizards, its colors, and its possibility of winning or losing, is now presented by subsets and their complements of this sample space, and so for the other lizard. Concretely, this means that we have $L_o \subseteq \Omega$, $L_o^C \subseteq \Omega$, representing the first orange morph, with $\mu(L_o)$ and $\mu(L_o^C)=1-\mu(L_o)$ the probabilities of winning or losing in a competition, respectively . Similarly we have $L_y \subseteq \Omega$, $L_y^C \subseteq \Omega$, representing the first yellow morph, with $\mu(L_y)$ and $\mu(L_y^C)=1-\mu(L_y)$ the probabilities of respectively winning or losing in a competition, and $L_b \subseteq \Omega$, $L_b^C \subseteq \Omega$, representing the first blue morph, with $\mu(L_b)$ and $\mu(L_b^C)=1-\mu(L_b)$ the probabilities of respectively winning or losing in a competition. We have an equivalent situation in the sample space for the second morph, hence $M_o \subseteq \Omega$, $M_o^C \subseteq \Omega$, representing the second orange morph, with $\mu(M_o)$ and $\mu(M_o^C)=1-\mu(M_o)$ the probabilities of respectively winning or losing in a competition. Similarly, we have $M_y \subseteq \Omega$, $M_y^C \subseteq \Omega$, representing the second yellow morph, with $\mu(M_y)$ and $\mu(M_y^C)=1-\mu(M_y)$ the probabilities of winning or losing in a competition, respectively, and $M_b \subseteq \Omega$, $M_b^C \subseteq \Omega$, representing the second blue morph, with $\mu(M_b)$ and $\mu(M_b^C)=1-\mu(M_b)$ the probabilities of winning or losing in a competition, respectively .

For the situation of both morphs joining in a competition, we consider the joint probabilities as constructed in a Kolmogorovian approach. Hence, for example, $\mu(L_o \cap M_y)$ represents the probability that the first morph being orange and the second being yellow, both win, $\mu(L_o \cap M_y^C)$ represents the probability that the first morph being orange wins and the second being yellow loses, $\mu(L_o^C \cap M_y)$ represents the probability that the first morph being orange loses and the second being yellow wins, and $\mu(L_o^C \cap M_y^C)$ represents the probability that the first morph being orange and the second being yellow both lose.

It is easy to show that the marginal law is always satisfied within such a Kolmogorovian model. Indeed, we have
\begin{eqnarray}
\mu(L_y)&=&\mu(L_y \cap (M_o\cup M_o^C))=\mu((L_y\cap M_o)\cup(L_y\cap M_o^C) \nonumber \\
&=&\mu(L_y\cap M_o)+\mu(L_y\cap M_o^C)
\end{eqnarray}
but also
\begin{eqnarray}
\mu(L_y)&=&\mu(L_y \cap (M_b\cup M_b^C))=\mu((L_y\cap M_b)\cup(L_y\cap M_b^C) \nonumber \\
&=&=\mu(L_y\cap M_b)+\mu(L_y\cap M_b^C)
\end{eqnarray}
and also
\begin{eqnarray}
\mu(L_y)&=&\mu(L_y \cap (M_y\cup M_y^C))=\mu((L_y\cap M_y)\cup(L_y\cap M_y^C) \nonumber \\
&=&\mu(L_y\cap M_y)+\mu(L_y\cap M_y^C)
\end{eqnarray} 
While from (\ref{marginal}), (\ref{marginalagain}) and (\ref{marginalagainn}), we know that these are not equal following from the experimental data.

\section{Conclusions\label{conclusions}}
Cyclic competition, an evolutionary analogue of the RPS game, is relevant to population ecology. However, modeling the intrinsic probabilities involved in RPS structures, revealed a fundamental difficulty. If the competitive encounters of the individual players are analyzed as subentities interacting within a compound entity, the joint probabilities describing the interactions turn out to violate Bell's inequalities and the marginal probability law. This implies that such an interaction cannot be modeled in a Kolmogorovian probability space where the joint probabilities are measures of conjunctions of the events of the individual players. Following typical investigations of similar situations in the foundations of quantum physics, it is known that an occurrence of violation of Bell's inequality and the marginal probability law is indicative of the presence of contextuality of a quantum nature. This means that the complex Hilbert space formalism of quantum mechanics is a natural candidate for the modeling of such a situation.

In this paper, we have worked with the set of experimental data collected in \cite{BleaySinervo2007} on the RPS cycles of the three colored morphs of the side-blotched lizard {\it Uta Stansburiana}. We have calculated in Section \ref{outcomeprobabilities} the joint probabilities from the data, and shown that they give rise to a non-ideal probabilistic RPS type of interaction dynamics for the competing lizard morphs. We have analyzed the contextuality following from this dynamics for the lizard morph encounters, and shown in Section \ref{contextuality} that indeed Bell's inequality and the marginal probability law are violated. In Section \ref{hilbertspacemodel} we have constructed an explicit complex Hilbert space description modeling this contextually and also in a faithful way all the experimental data, and we have proven that the operators representing the measurable quantities do not commute, which explicitly shows the presence of non-classicality in the situation. In Section \ref{compoundness} we have investigated in detail the structure of the lizard interaction making use of the Hilbert space representation developed in Section \ref{hilbertspacemodel}, and we have proven the presence of quantum entanglement in the situation of the competing lizard morphs. In this entanglement analysis we have made use of results of similarly studied situations in human cognition where quantum entanglement was encountered \citep{1asQI2013,2asQI2013,asIQSA2012}. Finally, we have identified in Section \ref{kolmogorov} why a single Kolmogorovian probability space, with joint probabilities as measures of interactions of individual events, is not possible.

The result may have far-reaching consequences, since several biological systems are believed to contain sub-dynamics of the cyclic competition type \citep{SinervoCalsbeek2006}. For example, the coexistence of a large number of phytoplankton species competing for a limited variety of resources in aquatic ecosystems (the paradox of the plankton) is believed to result from cyclic competition \citep{HW99,Huismanetal2001,Schippers}, and hence could incorporate the type of quantum-structural aspects we identified for the lizard morph interactions.

Our investigation links up with the new developments of `identification of quantum structure in domains different from the micro-world', and extends to animal behavior the results obtained in the context of human cognition. Furthermore, it suggests that ecological systems are intrinsically contextual, and constitutes a powerful support for systematic applications of quantum-structural modeling in biology.

\section*{Acknowledgments}
This research was supported by Grants G.0234.08 and G.0405.08 of the Flemish Fund for Scientific Research, and by NSF awards DEB8919600, DEB9307999, IBN9631757, IBN9629793, DEB0108577, IBN0213179, DEB0515973, DEB0918268 and IOS1022031 to B.S. and by the Arbelbide family, who generously provided access to their land.

\end{document}